\documentclass[12pt]{article}

\usepackage{amsmath,amssymb}

\newcommand{\beq}{\begin{eqnarray}}
\newcommand{\eeq}{\end{eqnarray}}

\begin{document}

\thispagestyle{empty}

\begin{center}

%\hspace{8cm}

\begin{flushright}
NSF-KITP-07-57\\
BCCUNY-HEP/07-03
\end{flushright}

\vspace{15pt}
{\large \bf GLUEBALL MASSES IN 
$(2+1)$-DIMENSIONAL ANISOTROPIC WEAKLY-COUPLED YANG-MILLS THEORY  }

\vspace{20pt}

{\bf Peter Orland}$^{\rm a.b.c.}$\!\footnote{giantswing@gursey.baruch.cuny.edu}

\vspace{8pt}

\begin{flushleft}
a. Kavli Institute for Theoretical Physics, The University of California, Santa Barbara, CA 
93106, U.S.A.
\end{flushleft}
 
\begin{flushleft}
b. Physics Program, The Graduate School and University Center,
The City University of New York, 365 Fifth Avenue,
New York, NY 10016, U.S.A.
\end{flushleft}

\begin{flushleft}
c. Department of Natural Sciences, Baruch College, The 
City University of New York, 17 Lexington Avenue, New 
York, NY 10010, U.S.A. 
\end{flushleft}

\vspace{40pt}

{\bf Abstract}
\end{center}

\noindent 
The confinement problem has been solved
in the anisotropic
$(2+1)$-dimensional SU($N$) Yang-Mills theory at weak coupling. In
this paper, we find the low-lying spectrum for $N=2$. The
lightest excitations are pairs of fundamental particles of the $(1+1)$-dimensional
${\rm SU}(2) \times {\rm SU}(2)$ principal chiral nonlinear
sigma model 
bound in a linear potential, with a specified matching condition where the
particles overlap. This
matching condition can be determined from the exactly-known
S-matrix for the sigma model. 

\hfill

\newpage

\setcounter{footnote}{0}

\setcounter{page}{1}

\section{Introduction}
\setcounter{equation}{0}
\renewcommand{\theequation}{1.\arabic{equation}}

In recent papers, some new techniques have been developed for calculating quantities in a
$(2+1)$-dimensional SU($N$)
gauge theories \cite{PhysRevD71}, \cite{PhysRevD74}, \cite{PhysRevD75}. These techniques
exploit the fact that in an anisotropic limit of small coupling, the gauge theory becomes
a collection of completely-integrable quantum field theories, namely 
${\rm SU}(N) \times {\rm SU}(N)$ principal chiral nonlinear
sigma models. These integrable
systems are decoupled, save for a constraint which is necessary for complete
gauge invariance. In the case of $N=2$, is possible to perturb away from integrability, using 
exactly-known off-shell matrix elements of the integrable theory.

Though the gauge theory
we consider s not spatially-rotation invariant, it 
has features one expects of real $(3+1)$-dimensional QCD; it is
asymptotically free and confines quarks at weak coupling. Thus the limit of no
regularization is accessible. 

One can formally remove the regulator in strong-coupling
expansions of $(2+1)$-dimensional gauge theories; the vacuum state in
this expansion yields a string tension and a mass gap which have
formal continuum limits. This can be done in a Hamiltonian lattice formalism \cite{Greensite}, 
or with an ingenious choice of degrees of freedom and point-splitting regularization \cite{kar-nair}. This leaves open the question of whether these expressions
can be trusted at weak coupling (more discussion of this issue can be found in 
the introduction of reference
\cite{PhysRevD74}). In particular, one would like to rule out a deconfinement transition, or
different dependence of physical quantities on the coupling (as in compact QED
\cite{Polyakov}). There is a proposal for the vacuum state \cite{leigh-min-yel}, in
the formulation of reference \cite{kar-nair} which
seems to give correct values for some glueball masses \cite{MeyerTeper}, but this proposal 
evidently requires more mathematical justification.

In this paper, we will work out the masses of the lightest glueballs for the case of gauge
group SU($2$). Our method would also work in principle for SU($N$) gauge theories, and our
reason for choosing $N=2$ is that the analysis is simplest for that case. 

The basic connection between the gauge theory and integrable systems is most
easily seen in axial gauge \cite{PhysRevD71}. The string tensions in 
the $x^{1}$-direction and $x^{2}$-direction (which we called the
horizontal and vertical string tensions, respectively)
for very small $g_{0}^{\prime}$, were found by simple physical
arguments. The result for the horizontal
string tension was confirmed for gauge group SU($2$), and additional corrections in
$g_{0}^{\prime}$ were found \cite{PhysRevD74}, through the use of
exact form factors for the currents of the sigma model. String tensions for higher
representations can also be worked out, and adjoint sources are not
confined  \cite{PhysRevD75}.

Careful derivations of the connection between the gauge theory and integrable systems
use the Kogut-Susskind lattice formalism \cite{PhysRevD71}, \cite{PhysRevD74}. A shorter derivation was given in reference \cite{Conf7}, which we summarize again here. The formalism is essentially
that of  ``deconstruction" \cite{deconstruction}. 

The Yang-Mills action is $\int d^{3} {\mathcal L}$, where the Lagrangian is
%\beq
${\mathcal L}=\frac{1}{2e^{\prime\,\,2}} {\rm Tr} F_{01}^{2} +
\frac{1}{2e^{2}} {\rm Tr} F_{02}^{2}
-\frac{1}{2e^{2}} {\rm Tr} F_{12}^{2}$,
%\label{YMlagrangian}
%\eeq
and where  
$A_{0}$,$ A_{1}$ and $A_{1}$ are SU($N$)-Lie-algebra-valued components of the gauge field,
and the
field strength is  $F_{\mu\nu}=\partial_{\mu}A_{\nu}-\partial_{\nu}A_{\mu}-i[A_{\mu},A_{\nu}]$. This
action is invariant under the
gauge transformation 
$A_{\mu}(x) \rightarrow ig(x)^{-1}[\partial_{\mu}-iA_{\mu}(x)]g(x)$, where $g(x)$ is an SU($N$)-valued
scalar field. We take $e^{\prime}\neq e$, thereby losing rotation invariance.

We discretize the $2$-direction, so that the $x^{2}$ takes on the
values $x^{2}=a,2a,3a\dots$, where
$a$ is a lattice spacing. All fields are considered functions of $x=(x^{0},x^{1}, x^{2})$. We define
the unit vector ${\hat 2}=(0,0,1)$. We 
replace $A_{2}(x)$ by a field $U(x)$ lying in
SU($N$), via $U(x)\approx \exp -iaA_{2}(x)$. There is a natural discrete covariant-derivative 
operator:
%\beq
${\mathcal D}_{\mu}  {\mathfrak U}(x)
=\partial_{\mu}{\mathfrak U}(x)-{\rm i}A_{\mu}(x){\mathfrak U}(x)+{\rm i}{\mathfrak U}(x)A_{\mu}(x+{\hat 2}a)$,
%\;, \nonumber 
%\eeq
$\mu=0,1$, for any $N\times N$ complex matrix field ${\mathfrak U}(x)$. The action is
$S=\int dx^{0} \int dx^{1}\sum_{x^{2}}\;a\;{\mathcal L}$ where
\beq
{\mathcal L}
%\!&\!\!=\!\!&\! 
=\frac{1}{2(g_{0}^{\prime})^{2}a} {\rm Tr} F_{01}^{2} +
\frac{1}{2g_{0}^{2}}{\rm Tr} [{\mathcal D}_{0} U(x)]^{\dagger}
{\mathcal D}_{0}U(x) 
%\nonumber \\
%&\!\!-\!\!&
-\frac{1}{2g_{0}^{2}}{\rm Tr} [{\mathcal D}_{1} U(x)]^{\dagger}{\mathcal D}_{1}U(x) \;,
\label{sigmalagrangian}
\eeq
and where $g_{0}^{2}=e_{0}^{2}a$ and $(g_{0}^{\prime})^{2}=e^{\prime\,\, 2}a$. The Lagrangian 
(\ref{sigmalagrangian}) is invariant under the gauge transformation:
%\beq
$A_{\mu}(x) \rightarrow ig(x)^{-1}[\partial_{\mu}-iA_{\mu}(x)]g(x)$ and
$U(x)\rightarrow g(x)^{-1}U(x)g(x+{\hat 2}a)$
%\label{sigma-cont-gauge-trans}
%\eeq
where again, $g(x)\in {\rm SU}(N)$ and $\mu$ is restricted to $0$ or $1$. The bare
coupling constants $g_{0}$ and
$g_{0}^{\prime}$ are dimensionless. We recover from (\ref{sigmalagrangian}) 
the anisotropic continuum action in the 
limit $a\rightarrow 0$. The sigma model field
is $U(x^{0},x^{1},x^{2})$, and each discrete $x^{2}$ corresponds to a different sigma 
model. The system (\ref{sigmalagrangian}) is 
a collection of parallel $(1+1)$-dimensional ${\rm SU}(N)\times {\rm SU}(N)$
sigma models, each of which couples to the auxiliary fields $A_{0}$, $A_{1}$. The 
sigma-model self-interaction is the dimensionless number
$g_{0}$. 

We feel it worth commenting on the nature of the anisotropic regime and how
it is different from the standard $(2+1)$-dimensional Yang-Mills theory. The point
where the regulator can be removed in the theory is $g_{0}^{\prime}=g_{0}=0$. This point
can be reached in our treatment, but only if
\beq
(g_{0}^{\prime})^{2} \ll \frac{1}{g_{0}}e^{-4\pi/(g_{0}^{2}N)}\;. \label{relative-scales}
\eeq
The left-hand side and ride-hand side are proportional to the two
energy scales in the theory (the latter comes from the two-loop beta function of
the sigma model). Thus our method cannot accommodate fixing the ratio
$g_{0}^{\prime}/g_{0}$,
which is natural in standard perturbation theory \cite{standardanisotropic}. This is
why the mass gap is not of order $e$, $e^{\prime}$ and the string tension is
not of order $e^{2}$, $(e^{\prime})^{2}$.

We now discuss the Hamiltonian in the axial gauge $A_{1}=0$. The left-handed and right-handed currents are, 
$j^{\rm L}_{\mu}(x)_{b}={\rm i}{\rm Tr}\,t_{b} \, \partial_{\mu}U(x)\, U(x)^{\dagger}$ and
$j^{\rm R}_{\mu}(x)_{b}={\rm i}{\rm Tr}\,t_{b} \, U(x)^{\dagger}\partial_{\mu}U(x)$, respectively, 
where $\mu=0,1$. The Hamiltonian obtained from (\ref{sigmalagrangian}) is $H_{0}+H_{1}$, where
\beq
H_{0}\!=\!\sum_{x^{2}}\int dx^{1} \frac{1}{2g_{0}^{2}}\{ [j^{\rm L}_{0}(x)_{b}]^{2}+[j^{\rm L}_{1}(x)_{b}]^{2}\}
%\!=\!\int dx^{1} \frac{1}{2g_{0}^{2}}\{ [j^{\rm R}_{0}(x)_{b}]^{2}+[j^{\rm R}_{1}(x)_{b}]^{2}\}
\;,\label{HNLSM}
\eeq
and
\beq
H_{1}\!\!&\!\!=\!\!&\!\! \sum_{x^{2}}  \int dx^{1} \,
\frac{(g_{0}^{\prime})^{2}a^{2}}{4}\,\partial_{1}\Phi(x^{1},x^{2})\partial_{1}\Phi(x^{1},x^{2}) \nonumber \\
\!\!&\!\!-\!\!&\!\! 
\left(\frac{g_{0}^{\prime}}{g_{0}}\right)^{2}\,\,\sum_{x^{2}=0}^{L^{2}-a}  \int dx^{1} \!\!
\left[ j^{\rm L}_{0}(x^{1},x^{2})\Phi(x^{1},x^{2}) -j^{\rm R}_{0}(x^{1},x^{2}) \Phi(x^{1},x^{2}+a) \right]  
\nonumber \\
&+&\;\;\;\;\;(g_{0}^{\prime})^{2}q_{b}\Phi(u^{1},u^{2})_{b} -(g_{0}^{\prime})^{2}
q^{\prime}_{b}\Phi(v^{1},v^{2})_{b}   \; ,
\label{continuum-local}
\eeq
where $-\Phi_{b}=A_{0\,\,b}$ is the temporal gauge field, and
where in the last term
we have inserted two color charges - a quark with charge $q$ at site $u$
and an anti-quark with charge $q^{\prime}$ at site $v$. Some gauge invariance remains
after the axial-gauge fixing, namely that 
for each $x^{2}$
\beq
\left\{ \int d x^{1}\left[ j^{L}_{0}(x^{1},x^{2})_{b}-j^{R}_{0}(x^{1},x^{2}-a)_{b}\right] - g_{0}^{2}Q(x^{2})_{b} \right\}\Psi=0\;,
\label{physical}
\eeq
where $Q(x^{2})_{b}$ is the total color charge from quarks at $x^{2}$ and $\Psi$ is any physical 
state. To derive the constraint (\ref{physical}) more precisely, we started with open boundary
conditions in the $1$-direction and periodic boundary conditions in
the $2$-direction, meaning that the two-dimensional space is a cylinder
\cite{PhysRevD71}, \cite{PhysRevD74}.

From (\ref{continuum-local}) we see that the left-handed charge of the sigma model
at $x^{2}$ is coupled to the electrostatic potential $\Phi$, at $x^{2}$. The right-handed charge
of the sigma model is coupled to the electrostatic potential at $x^{2}+a$. The 
excitations of $H_{0}$, which we call Fadeev-Zamoldochikov or FZ particles, behave like solitons, though
they do not correspond to classical configurations. Some of these FZ particles 
are elementary and others are bound states of
the elementary FZ particles. An elementary FZ particle has an adjoint charge and mass $m_{1}$. An 
elementary one-FZ-particle state
is a superposition of color-dipole states, with a quark  (anti-quark)
charge at $x^{1}, x^{2}$ and an anti-quark (quark)
charge at $x^{1},x^{2}+a$.  The interaction
$H_{1}$ produces a linear potential between color charges with the same value of $x^{2}$. Residual gauge
invariance (\ref{physical}) requires that at each value of $x^{2}$, the total color charge is zero. If there are 
no quarks with coordinate $x^{2}$, the total right-handed charge of FZ particles in the sigma model
at $x^{2}-a$ is equal to the total left-handed charge of FZ particles in the sigma model at $x^{2}$.

The particles of the principal chiral sigma model carry a quantum number 
$r$, with the values $r=1,\dots,N-1$
\cite{abda-wieg}.  Each particle of label $r$ has an antiparticle 
of the same mass, with label $N-r$. The
masses are given by
\beq
m_{r}=m_{1}\frac{\sin\frac{r\pi}{N}}{\sin\frac{\pi}{N}},\;\; m_{1}=K\Lambda(g_{0}^{2}N)^{-1/2}e^{-\frac{4\pi}{g_{0}^{2}N}}+{\rm non\!-\! universal \;corrections}\;, \label{mass-spectrum}
\eeq
where $K$ is a non-universal constant and $\Lambda$ is the ultraviolet cut-off
of the sigma model.
 
Lorentz invariance
in each $x^{0},x^{1}$ plane is manifest. For this reason, the 
linear potential is not the only effect of
$H_{1}$. The interaction creates and destroys pairs of elementary
FZ particles. This effect is quite small, provided that $g_{0}^{\prime}$ is small 
enough. Specifically, this means that the square of
the $1+1$ string tension in the $x^{1}$-direction coming from $H_{1}$ is small compared to
the square of the mass of fundamental FZ particle; this is just the condition
(\ref{relative-scales}). The effect is important, however, in that it is responsible for the correction to
the horizontal string discussed in the 
next paragraph in equation (\ref{string-tension}).

Simple arguments readily show that at leading order in $g_{0}^{\prime}$, the vertical and
horizontal string tensions are given by
\beq
\sigma_{\rm V}=\frac{m_{1}}{a}\;,\;\; \sigma_{\rm H}=\frac{(g_{0}^{\prime})^{2}}{2a^{2}}C_{N}\;, \label{string-tensions}
\eeq
respectively, where
$C_{N}$ is the smallest eigenvalue of the Casimir of 
${\rm SU}(N)$. 
These naive results for the string tension have further corrections in $g_{0}^{\prime}$, which were 
determined for the horizontal string tension for SU($2$) \cite{PhysRevD74}:
\beq
\sigma_{\rm H} =\frac{3}{2}\left( \frac{g_{0}^{\prime}}{a}\right)^{2} \left[ 1+
\frac{4}{3}\frac{0.7296}{K^{2}\pi^{2}}\Lambda a\frac{(g_{0}^{\prime})^{2}}{g_{0}^{2}}e^{4\pi/g_{0}^{2}} \right]^{-1}\;.
\label{string-tension}
\eeq
The leading term agrees with (\ref{string-tensions}). This calculation was done using the exact form factor for sigma model currents obtained by
Karowski and Weisz \cite{KarowskiWeisz}. The form factor can also be used to find corrections
of order $(g_{0}^{\prime})^{2}$ to the vertical string tension; this problem should be
solved soon. If the reader is not familiar with form-factor techniques in relativistic
integrable field theories, a self-contained review is in the appendix of reference 
\cite{PhysRevD74}.

Another recent application of exact form factors to the
$(2+1)$-dimensional SU($2$) gauge theory 
is reference \cite{CaselleGrinzaMagnoli}, in which
form factors of the two-dimensional Ising model \cite{McCoy} are used to
find the profile of the electric string near the high-temperature deconfining 
transition, assuming the Svetitsky-Yaffe conjecture \cite{SvetitskyYaffe}.

A rough picture of a gauge-invariant state for the gauge group SU($2$) with no quarks
is given in Figure 1. For $N>2$, there
are more complicated ways in which strings can join particles. For example, a junction of
$N$ strings is possible. Figure 1 is inaccurate in an important respect; the ``ring" of particles held together
by horizontal strings is extremely broad in extent in the $x^{2}$-direction compared
to the $x^{1}$-direction. This is because $\sigma_{\rm H}\ll \sigma_{\rm V}$.

The lightest states have
the smallest number of particles, by virtue of $\sigma_{\rm H}\ll \sigma_{\rm V}$. Thus 
the lightest glueballs are pairs of 
FZ particles with the same value of $x^{2}$. For 
small enough $g_{0}^{\prime}$, the very lightest state has a mass well-approximated
by $2m_{1}$. The purpose of this paper is to find the leading corrections in $(g_{0}^{\prime})^{2}$
to this result. This will be done using the S-matrix of the sigma model 
and the WKB formula. There are further small corrections, due to the softening of the
potential near where particles overlap, which we do do not determine.

It is clear that the lightest bound states of FZ particles are $(1+1)$-dimensional in character. If
we formulated a gauge theory in which $x^{2}$ was fixed in $U(x^{0},x^{1},x^{2})$, we would
find the same spectrum, as a function of
$m_{1}$ and $\sigma_{\rm H}$. In the Kogut-Susskind lattice formulation, a long row of plaquettes with open boundary
conditions is a regularized gauge theory of this type. The only real difference 
between this $(1+1)$ dimensional
model and that we study is that $\sigma_{\rm H}$ will receive different corrections of order
$(g_{0}^{\prime})^{2}$.

\begin{center}

\begin{picture}(150,60)(30,0)

\linethickness{0.5mm}

\multiput(75,-0.5)(0,7){8}{\multiput(0,0)(5,0){12}{\put(0,0){$-$}}}

\put(95,11.5){\circle*{7}}
\put(85,18.5){\circle*{7}}
\put(110,25.5){\circle*{7}}
\put(99,32.5){\circle*{7}}
\put(115,11.5){\circle*{7}}
\put(125,18.5){\circle*{7}}
\put(130,25.5){\circle*{7}}
\put(120,32.5){\circle*{7}}

\put(95,8.1){\line(1,0){20}}
\put(115,15.3){\line(1,0){10}}
\put(125,22.3){\line(1,0){5}}
\put(100,29.3){\line(1,0){10}}
\put(120,36.3){\line(-1,0){21}}
\put(129,29.3){\line(-1,0){9}}
\put(110,22.3){\line(-1,0){25}}
\put(85,15.3){\line(1,0){10}}

\end{picture}
\end{center}

\vspace{5pt}

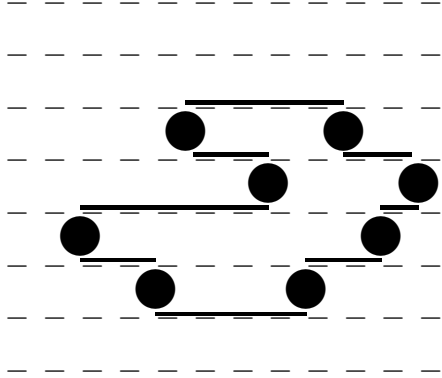
Figure 1. A glueball state is a collection of heavy particles, held weakly together

by strings. The horizontal coordinate is $x^{1}$ and the vertical coordinate is $x^{2}$.

\vspace{15pt}

In the next section we will discuss the wave function of an unbound pair of FZ
particles. We find that this is described by phase shift for the color-singlet sector. In Section
3, we determine the bound-state spectrum. The problem we solve is very similar to that
of two particle-states of the two-dimensional Ising model with an external magnetic
field \cite{McCoyWu} (for a good summary of
this problem, see reference \cite{BhaseenTsvelik}); the only genuine difference is the presence of a matching condition
where the particles overlap. This matching condition comes from the phase
shift of the scattering problem. We present our conclusions in Section 4.

\section{Scattering states of FZ particles}
\setcounter{equation}{0}
\renewcommand{\theequation}{2.\arabic{equation}}

The lightest glueball state, as discussed above, is simply a pair of FZ particles
located at the points $(x^{1},x^{2})$
and $(y^{1},x^{2})$ and bound in a linear potential. Residual
gauge
invariance (\ref{physical}), demands that the state be a color singlet. To begin with, however, we
simply write the form of a free state of two particles.

The state of the ${\rm SU}(2)\times {\rm SU}(2)
\simeq O(4)$ nonlinear sigma model with a particles of momenta $p_{1}$ and
$p_{2}$ and quantum numbers $j_{1}$ and $j_{2}$ (which take the values
$1,2,3,4$) is described by the wave function
\beq
\psi_{p_{1} p_{2}}(x^{1},y^{1})_{j_{1},j_{2}}
=\left\{ \begin{array}{cc}
e^{{\rm i} p_{1} x^{1}+{\rm i} p_{2} y^{1}} A_{j_{1},j_{2}}\;,\;& x^{1}<y^{1} \\
e^{{\rm i} p_{2} x^{1}+{\rm i}p_{1} y^{1}}\sum_{k_{1},k_{2}=1}^{4}S^{k_{1}k_{2}}_{j_{1}j_{2}}(p_{1},p_{2}) A_{k_{2},k_{1}}\;,\;& 
x^{1}>y^{1}
\end{array}
\right.\; , \label{free-two-particle}
\eeq
where $A_{j_{1}j_{2}}$ is an arbitrary set of complex numbers and
$S^{k_{1}k_{2}}_{j_{1}j_{2}}(p_{1},p_{2})$ is the two-particle S-matrix. We 
have not yet imposed (\ref{physical}).

The wave function (\ref{free-two-particle}) is written in a form where the $O(4)$ symmetry
is manifest. It is straightforward to write it in a form where the left ${\rm SU}(2)_{\rm L}$ and
the right ${\rm SU}(2)_{\rm R}$ symmetries are manifest, by writing
\beq
\psi_{p_{1} p_{2}}(x^{1},y^{1})_{a,{\bar b}}^{{\bar c},d}
=\sum_{j_{1},j_{2}} \frac{1}{\sqrt 2}(\delta^{j_{1} 4}_{ac}-{\rm i}\sigma^{j_{1}}_{ac})
\frac{1}{\sqrt 2}(\delta^{j_{2} 4}_{bd}-{\rm i}\sigma^{j_{2}}_{bd})^{*}\;
\psi_{p_{1} p_{2}}(x^{1},y^{1})_{j_{1},j_{2}} \label{SU(2)-freeparticle}
\eeq
describing a pair of color dipoles, one with quantum numbers $a,{\bar b}$ and 
the other with quantum numbers ${\bar c}, d$, where $\sigma^{j}$, $j=1,2,3$ denotes
the Pauli matrices.

We impose the physical state condition (\ref{physical}) on
(\ref{SU(2)-freeparticle}) by requiring that $a=b$ and $c=d$ and summing over
these colors. The projected wave function is, up to an overall constant,
\beq
\psi_{p_{1} p_{2}}(x^{1},y^{1})
=\left\{ \begin{array}{cc}
e^{{\rm i} p_{1} x^{1}+{\rm i}p_{2} y^{1}}\;,\;& x^{1}<y^{1} \\
e^{{\rm i} p_{2} x^{1}+{\rm i}p_{1} y^{1}}S_{0}(p_{1},p_{2}) \;,\;& 
x^{1}>y^{1}
\end{array}
\right.\; , \label{free-singlet-two-particle}
\eeq
where $S_{0}(p_{1},p_{2})$ is the singlet projection of the O($4$) S-matrix. This S-matrix
was first obtained by Zamolodchikov and Zamolodchikov \cite{Zamolodchikov}. A useful
form is given in reference \cite{KarowskiWeisz}:
\beq
S_{0}(p_{1},p_{2})=S_{0}(\theta)=-\frac{\pi-{\rm i}\theta}{\pi+{\rm i}\theta}
\exp {\rm i} \int_{0}^{\infty} \frac{d\xi}{\xi} \frac{1-e^{-\xi}}{1+e^{\xi}}
\sin \frac{\xi\theta}{\pi}\;, \label{singlet-S-matrix}
\eeq
where the relative rapidity $\theta$ is given by $\theta=\theta_{2}-\theta_{1}$, 
$p_{1}=m\sinh \theta_{1}$, $p_{2}=m\sinh \theta_{2}$ and where we denote the
particle mass $m_{1}$, given by (\ref{mass-spectrum}), by $m$ (because there is only one mass for the case of $N=2$). A derivation
of (\ref{singlet-S-matrix}) is in the appendix of reference \cite{PhysRevD74}.

The singlet S-matrix is just given by a phase shift $\phi(\theta)$:
$S_{0}(\theta)=\exp {\rm i} \phi(\theta)$. The phase shift has a simple form in the low-energy, non-relativistic limit, $\vert p_{1}-p_{2} \vert \ll m$. In this 
limit, $\theta\approx \vert p_{1}-p_{2}\vert/m$. The 
integral on the right-hand side of (\ref{singlet-S-matrix}) can be done by Taylor expanding
in $\vert p_{1}-p_{2}\vert/m$ yielding
\beq
\phi(\theta)=\phi(p_{1},p_{2})=\pi-\frac{3-2\ln 2}{\pi m}\vert p_{1}-p_{2} \vert\
+O\left(\frac{\vert p- r \vert^{2}}{m^{2}}\right)\;. \label{phase-shift} 
\eeq

\section{The low-lying glueball spectrum}
\setcounter{equation}{0}
\renewcommand{\theequation}{3.\arabic{equation}}

Let us now consider the states of a bound pair of FZ particles in the potential
$V(x^{1},y^{1})=2\sigma_{\rm H}\vert x^{1}-y^{1}\vert$
(the reason for the factor of two is simply that the particles are joined by a pair of strings). We use the non-relativistic approximation, used to find (\ref{phase-shift}). For our
problem, the horizontal string tension times the size of a typical bound state
is small compared to the mass, by (\ref{relative-scales}). This justifies the non-relativistic
approximation for low-lying states. The mass of a low-lying glueball is
given by
\beq
M=2m+E \;, \nonumber
\eeq
where $E$ is the energy eigenstate of the two-particle problem.

Let us introduce center-of-mass coordinates, $X=(x^{1}+y^{1})/2$ and $x=y^{1}-x^{1}$. The reduced
mass of the system is $m/2$. We factor out the phase depending on $X$, leaving us
only with a wave function depending on $x$. The Schr\"{o}dinger equation we consider is
\beq
-\frac{1}{m}\frac{d^{2}\psi}{dx^{2}}+2\sigma_{\rm H}\vert x \vert \psi=E\psi \nonumber
\eeq
with a matching condition at $x=0$ between the wave function $\psi(x)$ at $x>0$ and the
wave function at $x<0$. There is actually a further complication, which we do not
consider here; the potential changes slightly in the region where $x\approx 0$. This
is due to the fact that the color charge is slightly smeared out. This smearing out
can be calculated from the form factor \cite{KarowskiWeisz}.

Our results (\ref{free-singlet-two-particle}), (\ref{phase-shift}) for the unbound two-particle state,
tell us that for $x^{1}\approx y^{1}$, where the effect of the potential
can be ignored, the bound-state wave function in
the center-of-mass frame will
be of the form
\beq
\psi(x)
=\left\{ \begin{array}{cc}
\cos (px+\omega) \;,\;& x<0 \\
\cos [-px +\omega-\phi(p)] \;,\;& 
x>0
\end{array}
\right.\; , \label{bound-singlet-two-particle}
\eeq
for some angle $\omega$, where
$p=p_{1}-p_{2}$ and $\phi(p)=\pi-\frac{3-2\ln 2}{\pi m}\vert p\vert
+O( \vert p \vert^{2}/m^{2})$. The value
of $p$ near $x=0$ is given by $p=(mE)^{1/2}$, where $E$ is the energy eigenvalue of
the state. This is the matching condition between the wave function for $x>0$ and
for $x<0$.

The wave function for $x<0$ an Airy function. So is the wave function for $x>0$. We
therefore obtain the approximate WKB form
\beq
\psi(x)
=\left\{ \begin{array}{cc}
C(x+\frac{E}{2\sigma_{\rm H}})^{-1/4} \cos \left[ 
\frac{2}{3}(2m\sigma_{\rm H})^{1/2}(x+\frac{E}{2\sigma_{\rm H}})^{3/2}-\frac{\pi}{4} \right]
\;,\;& x<0 \\
\;C^{\prime} (\frac{E}{2\sigma_{\rm H}}-x)^{-1/4} \cos \left[ 
-\frac{2}{3}(2m\sigma_{\rm H})^{1/2}(\frac{E}{2\sigma_{\rm H}}-x)^{3/2}+\frac{\pi}{4} \right]\;,\;& 
x>0
\end{array}
\right.\; , \label{WKB}
\eeq
for some constants $C$ and $C^{\prime}$. The expression (\ref{WKB}) can be made to agree with
(\ref{bound-singlet-two-particle}) for small $x$, provided the generalization of the
Bohr-Sommerfeld quantization condition
\beq
\frac{2(m)^{1/2}}{3\sigma_{\rm H}} E_{n}^{3/2}
+\frac{3-2\ln 2}{\pi m^{1/2}} E_{n}^{1/2}-\left( n+\frac{1}{2}\right) \pi=0\;,\;\; n=0,1,2,\dots,
\label{Bohr-Sommer}
\eeq
is satisfied by $E=E_{n}$. The only new feature in this semi-classical formula is
the second term, produced by the phase shift. Absorbing the horizontal string
tension in the energy, by defining $u_{n}=E_{n}\sigma_{\rm H}^{-2/3}$, this cubic equation
becomes
\beq
\frac{2(m)^{1/2}}{3} u_{n}^{3/2}
+\frac{3-2\ln 2}{\pi m^{1/2}} \sigma_{\rm H}^{1/3}u_{n}^{1/2}-\left( n+\frac{1}{2}\right) \pi=0\;.\nonumber
\eeq
The second term can be ignored for sufficiently small $\sigma_{\rm H}$, i.e. sufficiently
small $g_{0}^{\prime}$.

There is a unique real solution of the cubic equation 
(\ref{Bohr-Sommer}) for a given integer $n \ge 0$, because $3-2\ln 2=1.613706>0$. The low-lying glueball masses are given by
\beq
M_{n}=2m+E_{n}=2m+\left[
\epsilon_{n}^{1/3}- \frac{3(3-2\ln2)\sigma_{\rm H} }{ 4\pi m} \epsilon_{n}^{-1/3}
\right]^{2}\;, \label{bound-state-spectrum}
\eeq
where
\beq
\epsilon_{n}=\frac{3\pi\sigma_{\rm H}( n+\frac{1}{2} ) }{4m^{1/2}}
+\left\{
\left[
\frac{3\pi\sigma_{\rm H}}{4m^{1/2}(n+\frac{1}{2})}
\right]^{2}
+\frac{1}{8}
\left[
\frac{3(3-2\ln2)\sigma_{\rm H}}{2\pi m}
\right]^{3}
\right\}^{1/2}.
\label{epsilon-definition}
\eeq

\section{Conclusions}

We have identified the low-lying glueballs of the anisotropic Yang-Mills
theory in $(2+1)$ dimensions as bound pairs of the fundamental massive particles
of the principal chiral nonlinear sigma model. We found a matching condition for the
bound-state wave function at the origin, which when combined with elementary
methods yields the spectrum of the lightest states.

There are other aspects of the two-particle bound-state problem we
have not considered here. First, the potential is not
precisely linear in the region where the two particles are close
together. The corrections to the potential
can be determined using form factors. This will 
slightly modify (\ref{Bohr-Sommer}). A completely
different issue is that there 
are small corrections to the form factors themselves, coming
from the presence of bound states. This, in turn, will give a further
correction to the horizontal string tension found in
\cite{PhysRevD74}. Such corrections
to form factors in theories close to integrability were first discussed by Delfino, 
Mussardo and Simonetti \cite{Delfino}.  The
bound-state energies proliferate between 
$2m$ and $4m$, as $g_{0}^{\prime}\rightarrow 0$. Our 
method breaks
down as the bound-state mass reaches $4m$, because the bound state develops an
instability towards fission into a pair of two-particle bound states. This is analogous to the
situation for the Ising model in a field \cite{McCoyWu}, \cite{BhaseenTsvelik}
as we stated earlier. It 
should be worthwhile to understand the relativistic corrections to the 
bound-state formula,
along the lines of the work 
of Fonseca and Zamolodchikov \cite{FonsecaZamolodchikov}. 

A similar calculation is possible for SU($N$). The exact S-matrix 
of the principal chiral nonlinear sigma model is 
known for $N>2$ \cite{abda-wieg}. An
interesting feature is that the phase shift should vanish as $N\rightarrow \infty$, 
with $g_{0}^{2}N$ fixed, meaning that the wave function would be
continuous where FZ particles overlap.

It would be interesting to study the scattering of a glueball by an external
particle. If the scattering is sufficiently short range, the FZ particles could be liberated from
the glueball, after which hadronization would ensue.

The results of this paper and of references
\cite{PhysRevD71} and \cite{PhysRevD74} may be extendable to 
the standard $(2+1)$-dimensional
isotropic Yang-Mills theory with $g_{0}^{\prime}=g_{0}$. The strategy we have in mind
is an 
anisotropic renormalization procedure. At the start is a standard field theory
with an isotropic cut-off. By anisotropically
integrating out high-momentum degrees
of freedom, the isotropic
theory will flow to an anisotropic theory with a small momentum
cut-off in the $x^{2}$-direction and a large momentum cut-off in the $x^{1}$ direction. If
the renormalized couplings satisfy the condition (\ref{relative-scales}), we could apply our
techniques. A check of such a method would be approximate
rotational invariance of the string tension. This would give an analytic 
first-principles method of solving the isotropic gauge theory with fixed
dimensionful coupling constant $e$, and no cut-off. The only
other analytic weak-coupling argument for a mass gap and
confinement in $(2+1)$-dimensions, namely that of orbit-space distance
estimates, discussed by Feynman \cite{feynman}, by Karabali and Nair in the
second of references \cite{kar-nair}, and by Semenoff and the
author \cite{orlandsemenoff} is suggestive, but
has not yielded 
definite results yet\footnote{See also reference \cite{metric} for a general discussion of
distance in orbit space.}.

\section*{Acknowledgments}

This research was supported in
part by the National Science Foundation under Grant No. PHY05-51164
and by a grant from the PSC-CUNY.

\end{document}